\documentclass{PoS}
\usepackage{amsmath,amssymb}

\newcommand{\preprint}{
  \begin{picture}(0,0)
    \put(0,120){{\rm\normalsize ADP-07-08/T648}}
  \end{picture}}

\title{\preprint%
Comparing SU(2) to SU(3) gluodynamics \linebreak[1] on large lattices}

\ShortTitle{Comparing SU(2) to SU(3) gluodynamics on large
  lattices}

\author{A. Sternbeck, L.\ von Smekal, D.\ B.\ Leinweber and 
        A.\ G.\ Williams\\
        Centre for the Subatomic Structure of Matter
        (CSSM), Department of Physics,\\
        University of Adelaide, SA 5005, Australia\\
        \mbox{Email: \email{andre.sternbeck@adelaide.edu.au},
               \email{lorenz.smekal@adelaide.edu.au},}
        \mbox{\email{dleinweb@physics.adelaide.edu.au},
              \email{anthony.williams@adelaide.edu.au}}}

\abstract{We study the SU(2) gluon and ghost propagators in Landau gauge on
  lattices up to a size of $112^4$. A comparison with the SU(3) case
  is made and finite-volume effects are then investigated. We find that
  for a large range of momenta the SU(2) and SU(3) propagators are
  remarkably alike. In the low-momentum region we compare with recent
  results obtained in DSE studies on a \mbox{4-torus}.}

\FullConference{The XXV International Symposium on Lattice Field Theory\\
		 July 30 -- 4 August 2007\\
		 Regensburg, Germany}

\newcommand{\Tr}{\operatorname{Tr}}                   

\newcommand{\Fig}[1]{Fig.~\ref{#1}}
\newcommand{\Tab}[1]{Table~\ref{#1}}

\begin{document}

\section{Introduction}

For quite some time now the infrared behaviour of the fundamental QCD
Green's functions has been a focus of intense study
(see, e.g., the reviews \cite{Alkofer:2000wg,Fischer:2006ub} and
references therein). Nonperturbative effects have been and continue 
to be investigated in parallel with various functional continuum
methods and in lattice simulations. The infrared behaviour of gluon
and quark propagators and vertex functions, especially in Landau
gauge, thereby is of particular relevance for an understanding of
confinement in the covariant formulation of QCD in terms of local
quark and gluon fields \cite{Alkofer:2000wg}. They also serve 
for model building in a hadron phenomenology based on these elementary QCD
correlations via Dyson-Schwinger and Bethe-Salpeter/relativistic
Faddeev equations \cite{Alkofer:2000wg,Roberts:1994dr,Roberts:2007jh}.

Looking back at an intensive decade of research, results obtained in
studies of Dyson-Schwin\-ger equations (DSEs) 
\cite{vonSmekal:1997isvonSmekal:1997vx,Lerche:2002ep}, within
stochastic quantisation \cite{Zwanziger:2001kw,Zwanziger:2003cf}, and
of the functional renormalisation group equations (FRGEs)
\cite{Pawlowski:2003hq,Pawlowski:2005xe} all agree in 
an infrared vanishing gluon propagator $D$ coupled to an enhanced
infrared-diverging ghost propagator $G$. In Landau gauge these are 
\begin{equation} D^{ab}_{\mu\nu}(q^2) =
\delta^{ab}\left(\delta_{\mu\nu} - \frac{q_{\mu}q_{\nu}}{q^2}
\right)\frac{Z(q^2)}{q^2}\qquad\textrm{and}\qquad G^{ab}(q^2) =
\delta^{ab}\frac{J(q^2)}{q^2}\;,
\end{equation} where the dressing functions $Z$ and $J$ are
predicted to follow power laws, namely
\begin{equation} Z(q^2) \propto
  (q^2)^{\kappa_D}\qquad\textrm{and}\qquad J(q^2)
\propto (q^{2})^{-\kappa_G}\qquad\textrm{for}\quad q^2\rightarrow0\;,
\end{equation} with exponents satisfying $\kappa_D = 2 \kappa_G$
\cite{vonSmekal:1997isvonSmekal:1997vx}. This infrared behaviour
which is determined by a single exponent $\kappa\equiv\kappa_G$
can be generalised to vertex functions with an arbitrary number of
ghost and gluon legs in a simple counting scheme for the pure gauge
theory \cite{Alkofer:2004it}. Moreover, comparing DSEs and FRGEs the
uniqueness of this behaviour has been shown in
Ref.~\cite{Fischer:2006vf}. Under the additional assumption that the
ghost-gluon vertex is finite and regular in the infrared, the value of
the infrared exponent is given by $\kappa \approx 0.596$
\cite{Lerche:2002ep,Zwanziger:2001kw,Pawlowski:2003hq}. 

Despite intensive efforts, however, this behaviour has so far not been
confirmed in simulations of lattice Landau-gauge QCD in 4 dimensions
(see, e.g., \cite{Bonnet:2001uhBowman:2004jm,Sternbeck:2006cg}). 
Rather the results of present lattice simulations are, more or less,
in favour of a non-vanishing 
gluon propagator and a diverging ghost propagator with a $\kappa_G$
value much less than that given above. The remaining discrepancy
between the functional approaches and lattice QCD results is quite 
unsatisfying and needs clarification.

One attempt at a better understanding of the discrepancy has been
undertaken in DSE studies on a finite torus. For recent results see
Refs.~\cite{Fischer:2007pf,Fischer:2007la}. There, 
qualitatively good agreement has been found when comparing the
momentum dependence of gluon and ghost propagators as found in
lattice QCD to the solutions from finite-volume DSEs (see, e.g., Fig.~8 in
Ref.~\cite{Fischer:2007pf}). In addition, these finite-volume solutions
approach the infinite-volume DSE results when increasing the volume.
Therefore, the discrepancy between the functional approaches and lattice QCD
results might be due to finite-volume effects. Also, recent
lattice results for an SU(2) gauge theory in 2 dimensions
\cite{Maas:2007uv} (on much larger 
lattices) are in quite compelling agreement with the corresponding
infrared behaviour as predicted by the continuum studies.

\bigskip

The objective of this study is to provide more information about 
the infrared behaviour of the Landau-gauge gluon and ghost propagators  
from a lattice-QCD perspective. For this, we first present numerical
evidence that towards the low-momentum region both propagators are
independent of whether considering SU(2) or SU(3).  Given this, we then
use the numerically cheaper gauge group SU(2) to dig even deeper into
the infrared region of lattice Landau-gauge QCD by using large
symmetric lattice volumes. First results of this study have been
presented previously~\cite{Trento07}. 

\begin{table}[t]
  \begin{minipage}[b]{0.29\linewidth}
  \centering
  \label{tab:stat}
  \caption{Number of configurations used for the different $\beta$ values
    and lattice sizes. Approximate values for the edge length $L$ are also
    given. For this we set $\sqrt{\sigma}=440\ \text{MeV}$ and use the
    $\sigma a^2$ results from Ref.~\cite{Langfeld:2007zw}.\vspace{-8.0ex}} 
 \end{minipage}
\quad\;
 \begin{minipage}[t]{0.6\linewidth}
  \centering
  \begin{tabular}{crrc@{\quad}|@{\quad}crcc}
    \hline\hline 
    $\beta$ & latt. & $L$ [fm] & \#conf. & $\beta$ & latt. & $L$ [fm]
    & \#conf.\\
    \hline
    2.3     & $16^4$ & 2.7 & 100 &  2.5 & $32^4$ & 2.8 & 100 \\
    2.3     & $32^4$ & 5.5 & 100 &  2.5 & $48^4$ & 4.3 & 50 \\
    2.3     & $56^4$ & 9.6 & 70  &  2.5 & $56^4$ & 5.0 & 70 \\
    2.3     & $80^4$ & 13.7 & 67  &  2.5 & $80^4$ & 7.1 & 23 \\
    2.3     & $112^4$ & 19.2 & 34  &  2.6 & $48^4$ & 3.1 & 100 \\
    \hline\hline
  \end{tabular}
 \end{minipage}
\end{table}

For our study we use the standard Wilson gauge action to generate
SU(2) gauge configurations for a couple of lattice sizes ranging from
$16^4$ up to $112^4$ at $\beta=2.3$, 2.5 and 2.6 (see \Tab{tab:stat}
for details). After every 2000
hybrid-overrelaxation updates, the gauge configurations are
gauge-fixed to Landau gauge using an overrelaxation algorithm. As
stopping criterion we chose
\begin{displaymath}
  \max_x\, \Tr\left[(\nabla_{\mu} A_{x,\mu})(\nabla_{\mu}
    A_{x,\mu})^{\dagger}\right] < 10^{-14}\;.
\end{displaymath}
$A_{x,\mu}$ are the lattice gluon fields given here in terms of
gauge-fixed links $U_{x,\mu}$ as
\begin{displaymath}
  A_{x,\mu}\equiv A_{\mu}(x+\hat{\mu}/2) =
  \frac{1}{2aig_0}(U_{x,\mu}-U_{x,\mu})\Big|_{\mathrm{traceless}} 
\end{displaymath}
On each such gauge-fixed configuration the momentum-space gluon and
ghost propagators are measured. On the lattice, the former is defined
as the Monte-Carlo average of the correlator
\begin{displaymath}
  D^{ab}_{\mu\nu}=\left\langle
    \tilde{A}^a_{\mu}(k)\tilde{A}^b_{\nu}(-k)\right\rangle_{U} 
\end{displaymath}
of Fourier-transformed gluon fields
$\tilde{A}_{\mu}=\tilde{A}^a_{\mu}T^a$. The ghost propagator 
can be estimated by
\begin{displaymath}
  G^{ab}(k) =
  \frac{1}{V}\left\langle\sum_{xy}\left(M^{-1}\right)^{ab}_{xy}\;e^{ik(x-y)}
  \right\rangle_U
\end{displaymath}
where $M$ is the lattice Faddeev-Popov operator in Landau gauge. For a
definition of $M$ and details on its inversion we refer to
\cite{Sternbeck:2005tk,Sternbeck:2006rd} and references therein.

\section{Comparing SU(2) to SU(3) results}

\begin{figure}[t]
  \centering
  \mbox{\includegraphics[height=7cm]{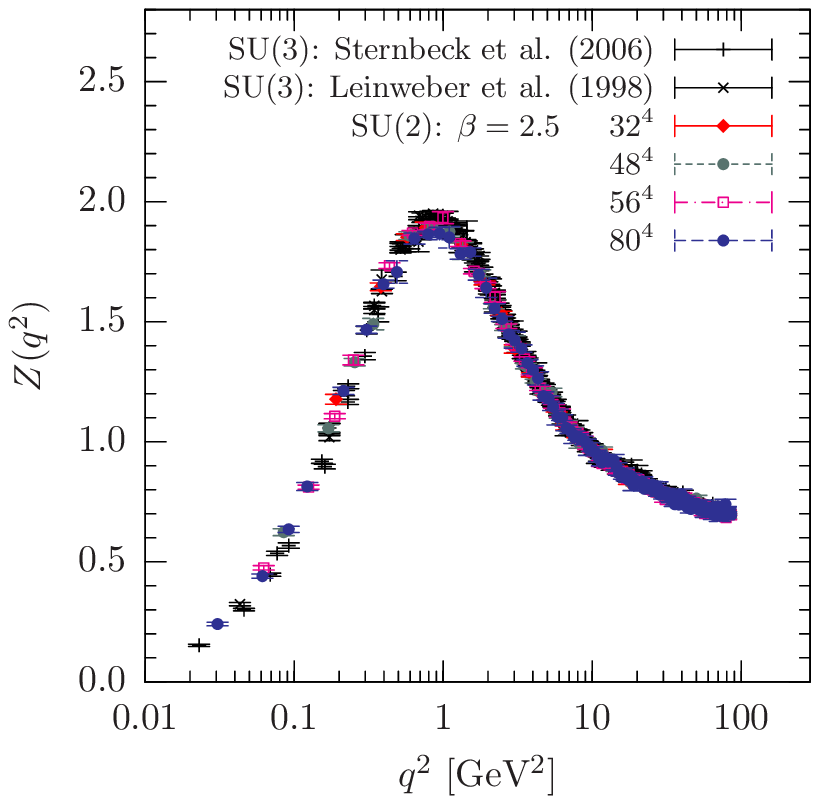}\qquad
        \includegraphics[height=7cm]{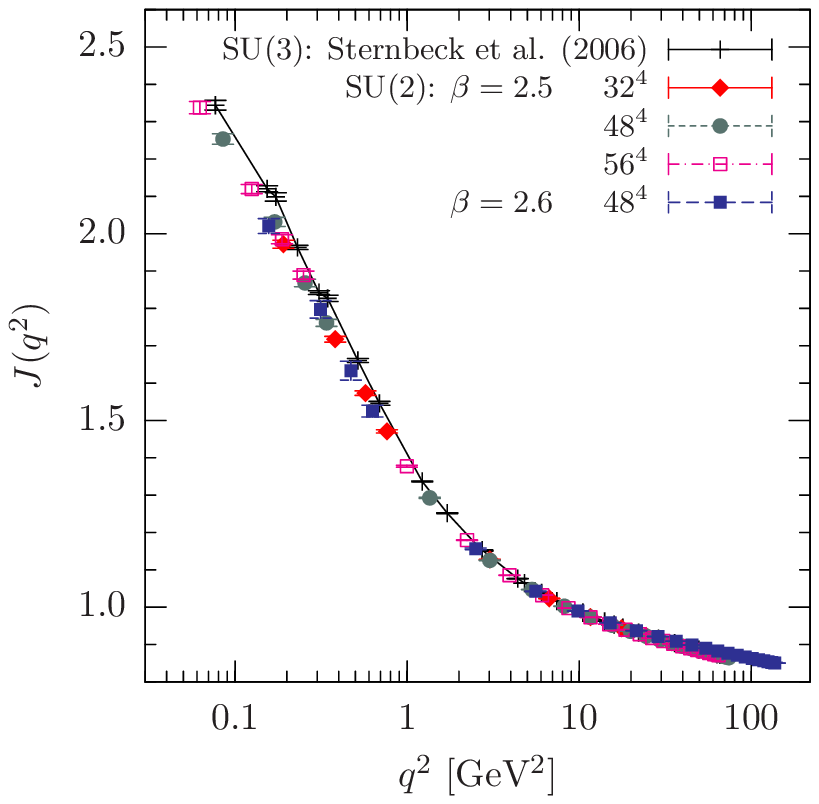}}
     \caption{The gluon (left) and ghost (right) dressing functions
       vs.\ $q^2$ for the gauge groups SU(2) (colour symbols) and SU(3)
       (black symbols). All data sets have been renormalised at
       $\mu=3$~GeV. The SU(3) ghost data are taken from
       Ref.~\cite{Sternbeck:2006cg}.\vspace{-0.4cm}}
     \label{fig:glgh_dress}  
\end{figure}

\medskip
\begin{floatingfigure}[r]
  \parbox{7.4cm}{%
    \;\includegraphics[width=7.8cm]{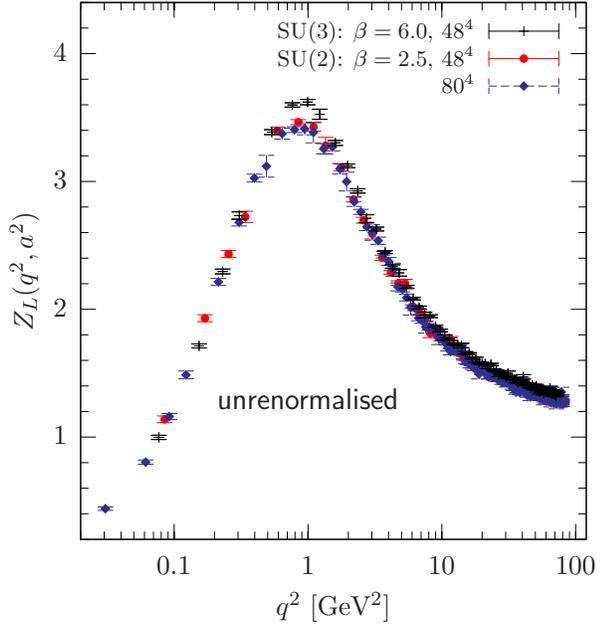}}
  \caption{The unrenormalized gluon dressing function as a function of
    $q^2$ (in physical units) for the gauge
    group SU(2) and SU(3).}
\label{fig:gl_dress_qq_su2su3_unren}
\end{floatingfigure}
Our results for the SU(2) gluon dressing function are shown in
Fig.~\ref{fig:glgh_dress} (left) together with corresponding SU(3) data taken
from Refs.~\cite{Sternbeck:2006cg,Leinweber:1998im}. Looking at those
figures, the data for the two gauge groups are remarkably alike almost
for the whole momentum region. A small deviation, however, is observed
around the hump which becomes more pronounced upon increasing the volume. 

The near $N_c$-independence becomes even more striking when comparing
\emph{un\-renorm\-alis\-ed data} at approximately equal lattice spacings and
volumes for both gau\-ge groups. Such a comparison is shown in
\Fig{fig:gl_dress_qq_su2su3_unren} where we compare our SU(2) data
for the gluon dressing function at {$\beta=2.5$} and with SU(3) data
at {$\beta=6.0$} \cite{Sternbeck:2006cg}. Obviously, 
the dependence on $N_c$ is rather small for a large range of momenta,
but it seems not to be that small around $q=1~\text{GeV}$. Also,
at higher momenta a small $N_c$-dependence is visible as expected from
perturbation theory given the $N_c$-dependence of the gluon dressing
function beyond 1-loop.\newline
\indent Considering next the ghost fields, the re\-normalised ghost
dressing functions of both gauge groups are compared in
\Fig{fig:glgh_dress} (right). As for the gluon propagator, the
dependence on the gauge group is rather small. This is also the case
for the unrenormalised data (not shown). A slight
deviation in the slope is visible towards low-lying momenta,
however. Whether this 
is a gauge-group rather than a Gribov-copy issue remains to be
investigated. Note that the Gribov ambiguity is expected to
introduce some systematic error, in particular for the ghost
propagator at lower momenta. For example, a bias towards larger values was
observed in Ref.~\cite{Sternbeck:2005tk} for momenta {$q<1$ GeV}. Since the
Landau-gauge condition for SU(2) is less ambiguous than for SU(3) the
slight deviation in the slope could be related
to this rather than to $N_c$. We leave a detailed investigation of
this for a future study.

\section{Studying finite-volume effects}

\begin{figure}[b]
\begin{minipage}[t]{0.48\linewidth}
  \centering
  \includegraphics[height=7.1cm]{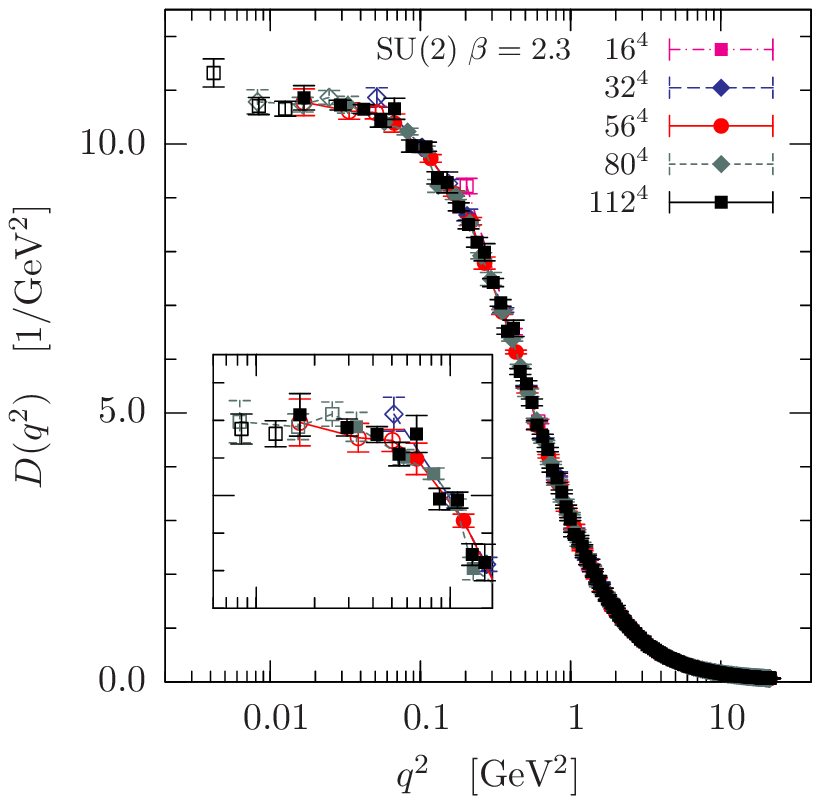}
  \caption{The unrenormalised SU(2) gluon propagator vs.\ $q^2$
    calculated using different lattice sizes at $\beta=2.3$. Data
    shown with open symbols would be subject to a cone cut.}
  \label{fig:gl_dress_b2p3}
\end{minipage}
\hfill
\begin{minipage}[t]{0.48\linewidth}
  \centering
  \includegraphics[height=7.1cm]{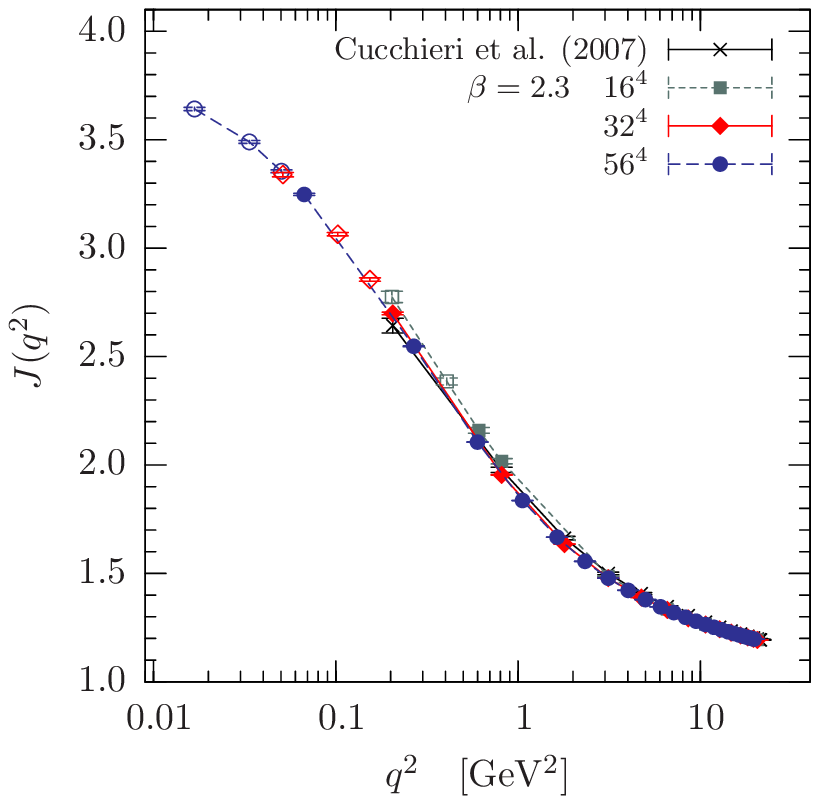}
  \caption{As in the left figure but for the SU(2) ghost dressing
    function. For comparison we include data from
    \cite{Cucchieri:2007ta}.}
  \label{fig:gh_dress_b2p3}
\end{minipage}
\end{figure}

Given the rather small $N_c$-dependence, a finite-volume study has
been performed for the numerically cheaper gauge group, SU(2). For this
the coupling constant has been fixed to $\beta=2.3$. Various
lattice sizes, ranging from $16^4$ to $112^4$, are considered. The
lattice sizes in physical units are listed in \Tab{tab:stat}.
Our data for the gluon propagator is shown in
\Fig{fig:gl_dress_b2p3} where data sets have neither
been renormalised nor cone cut.\footnote{\ A cone cut
  \cite{Leinweber:1998im} is usually imposed to reduce finite-volume
  effects. Such data is specially flagged with open symbols here and
  must be interpreted with due care. This should not be confused with
  a cylinder cut which we do apply to reduce discretisation errors.}
The statistics for the $112^4$ lattice is still rather limited.

From the figure we conclude that the finite volume mainly affects the data
only at the lowest (non-zero) lattice momentum. At larger momenta the
data agrees quite well, within errors, comparing different 
lattice sizes. This suggests that indeed a cone cut does well in
reducing finite-volume effects, but data at momenta like
$k=(1,1,1,0)$ (or permutations thereof) need not be excluded necessarily
from the analysis. 

Comparing to the DSE studies on a torus \cite{Fischer:2007pf}, the
finite-volume effect seen here agrees qualitatively with what has been
observed there. However, at the presently available volumes we cannot
confirm an infrared-decreasing gluon propagator. Note that according
to Ref.~\cite{Fischer:2007pf} these volumes are already in a range
where the gluon propagator is expected to decrease towards lower
momenta. Rather, our data suggests that the gluon propagator stays
finite in the zero-momentum limit. Similar results have been presented
at this conference by Cucchieri et al.\ \cite{Cucchieri:2007la}. See
also \cite{MMP:2007la}.

\medskip

The data for the SU(2) ghost dressing function for different volumes
is shown in \Fig{fig:gh_dress_b2p3}. There, a small volume
($16^4$) biases the smaller-momentum data towards larger
values. However, the dominant effect in the finite-volume DSE studies
is in the opposite direction. It will be interesting to consider
larger volumes to see if this effect persists. Calculations for the ghost 
propagator on a $80^4$ lattice are currently in progress and will be
reported elsewhere.

\section{Conclusions}

We have calculated SU(2) gluon and ghost propagators in Landau
gauge at different lattice spacings and volumes. A comparison of our
data with corresponding SU(3) results reveals that there is
very little $N_c$-dependence in the propagators over the whole
momentum range as first reported in \cite{Trento07}. In particular,
towards lower momenta, any signs for an $N_c$-dependence in the gluon
propagator disappear. Similar findings have been presented in
Refs.~\cite{Cucchieri:2007zmOrlando:2007la}. Moreover, the results found
here are consistent with an approximate overall $1/N_c$-scaling of the
nonperturbative running coupling constant 
\cite{vonSmekal:1997isvonSmekal:1997vx} 
\begin{displaymath}
  \alpha_s(p^2) = \frac{g^2_0(a)}{4\pi}\,Z_L(p^2,a^2)\,J_L^2(p^2,a^2)\;.
\end{displaymath}
Here $Z_L$ and $J_L$ are the lattice gluon and ghost dressing
functions, respectively. $g^2_0(a)$ is the coupling constant at the
cutoff scale $1/a$.
\footnote{Note that recently a lattice study has been commenced
  which considers this definition of $\alpha_s$ to estimate
  the $\Lambda$ parameter of QCD. First preliminary results have been
  presented at this conference \cite{Sternbeck:2007la}.}

Some small $N_c$-dependence is apparent in the gluon dressing function
around its hump and at large momenta. A slight variation in the ghost 
dressing function towards lower momenta may or may not be related to
$N_c$. Alternatively, this might be caused by the Gribov
ambiguity which is known to introduce a systematic uncertainty
here \cite{Sternbeck:2005tk}.

Our finite-volume study shows that the finite-lattice extent affects
the gluon propagator at the lowest non-zero lattice momentum beyond
the statistical error. The finite-volume effects qualitatively follow
those found in DSE studies on a torus
\cite{Fischer:2007pf,Fischer:2007la}. Of course, the volumes used at
present, especially for the ghost propagator, may not yet be large
enough to confirm the predicted finite-volume effects. However,
our data provides indication of a plateau in the gluon propagator for
momenta around and below approximately 200~MeV, or roughly
$\Lambda_{\textrm{QCD}}$, in contrast to the DSE results on a
torus. To see the onset of an infrared suppression at even smaller
momenta still requires a somewhat optimistic interpretation. Similar
findings for $SU(3)$ have been presented at this
conference~\cite{Bogolubsky:2007la}. It will be interesting to see
whether and how the results change, in particular at low momenta, when
using the slightly more expensive modified lattice Landau gauge of
\cite{vonSmekal:2007la} in comparably large volumes.

\section*{Acknowledgements}

This research was supported by the Australian Research Council.
Generous grants of time on the Hydra, Aquila and Corvus Supercomputers,
offered by the South Australian Partnership for Advanced Computing 
(SAPAC), allowed us to run simulations on such large lattice
sizes. Grants of time on the APAC facility for parts of the project
are also acknowledged.

\bibliographystyle{myunsrthep}
\bibliography{references}

\end{document}